\documentclass[
    ,final            % use final for the camera ready runs
  ]
  {aipproc}

\newcommand{\teffm}{$T_{\rm eff}$ }

\newcommand{\msunm}{$\rm M_\odot$}
\newcommand{\mstarm}{$\rm M_*$}

\newcommand{\logg}{$\log  \rm g$ }

\newcommand{\Mhem}{$\rm M_{He}$}
\newcommand{\Mhm}{$\rm M_{H}$}
\newcommand{\qfmm}{$\rm q_{fm}$}

\newcommand{\Xom}{$\rm X_{o}$}
\newcommand{\muhz}{$\mu$Hz}

\layoutstyle{8x11double}

\begin{document}

\title{Asteroseismological Analysis of Rich Pulsating White Dwarfs}

\classification{95.75.-z; 97.20.Rp}
\keywords      {White Dwarfs, Asteroseismology, G38-29, R808}

\author{A. Bischoff-Kim}{
  address={Department of Chemistry, Physics and Astronomy, CBX 82, Georgia 
College \& State University, Milledgeville, GA 31061}
}

%\author{<author2>}{
%  address={<common address for author2 and author3>}
%}

%\author{<author3>}{
%  address={<common address for author2 and author3>}
%  ,altaddress={<author1 address>} % additional visiting address
%}

\begin{abstract}
We present the results of the asteroseismological analysis of two rich
DAVs, G38-29 and R808, recent targets of the Whole Earth Telescope. 20
periods between 413~s and 1089~s were found in G38-29's pulsation
spectrum, while R808 is an even richer pulsator, with 24 periods
between 404~s and 1144~s. Traditionally, DAVs that have been analyzed
asteroseismologically have had less than half a dozen modes. Such a
large number of modes presents a special challenge to white dwarf
asteroseismology, but at the same time has the potential to yield a
detailed picture of the interior chemical make-up of DAVs. We explore
this possibility by varying the core profiles as well as the layer
masses. We use an iterative grid search approach to find best fit
models for G38-29 and R808 and comment on some of the intricacies of
fine grid searches in white dwarf asteroseismology. 
\end{abstract}

\maketitle

%%%%%%%%%%%%%%%%%%%%%%%%%%%%%%%%%%%%%%%%%%%%
%% MAINMATTER
%%%%%%%%%%%%%%%%%%%%%%%%%%%%%%%%%%%%%%%%%%%%

\section{Astrophysical context}

G38-29 and R808 are high amplitude, cool hydrogen dominated atmosphere 
pulsating white dwarfs (cDAVs). cDAVs are of interest for studying 
convection using the shape of their highly non-linear light curves
\citep{Montgomery07}. In addition, G38-29 and R808 are rich white dwarf 
pulsators, with over 20 periods present in their power spectrum. This makes 
them ideal subjects for asteroseismological studies. With the exception of 
hot pulsating white dwarfs \citep[e.g.][]{Corsico06a} , such rich pulsating 
white dwarfs have not been analyzed asteroseismologically before. Because of 
the large number of periods, asteroseismological identification of the modes
in these stars is necessary in order to keep the study of convection 
using non-linear light curve fitting techniques computationally tractable.

With such a large number of modes, we can begin to probe the interior
structure of white dwarfs in more detail. Cool pulsating white dwarfs 
have the advantage of being simpler to model than hot white dwarfs. The
chemical elements in their interiors are nearly fully settled and much of 
their interiors close to fully degenerate. In addition, there are over 100 
known DAVs \citep{Castanheira06} as opposed to 5 known DOVs \citep{Quirion07}.
Not all DAVs are rich pulsators, but with improved observational techniques, 
there is a potential to increase the number of modes observed in these 
stars, providing useful data for detailed asteroseismological analyses 
of white dwarf interiors.

In this paper, we present preliminary asteroseismological analyses of 
G38-29 and R808, based on Whole Earth Telescope (WET) campaigns performed
in fall 2007 and spring 2008 \citep{thompson09}. We are able to derive stellar 
parameters such as mass and effective temperature for each star as well as 
internal structure parameters. We suggest an identification for the modes
in each star.

\section{Spectroscopy and Power Spectra}

According to spectroscopy \citep{Beauchamp99}, G38-29 has a temperature 
of 11,180~K and a \logg of 7.91, which translates to a mass of 0.55 
\msunm. It was the object of a small WET campaign in November 2007. 
Analysis of the data revealed 20 independent modes, with periods ranging
between 413.307~s and 1089.39~s. Of special interest in the power spectrum
is an $\ell=1$ triplet, identified as such from average period spacing
arguments. The triplet is split by 7~\muhz, leading to a rotational 
period of 21 hrs for the star and providing ground for the identification
of additional multiplets (see Table \ref{t2}). Out of the 20 independent 
modes, 15 are candidate m=0 modes.

R808 is similar in temperature to G38-29 (11,160~K), but more massive 
with a \logg of 8.04, corresponding to a mass of 0.63 \msunm. It was the
object of a WET campaign in April 2008, where 25 independent modes were
found in its somewhat noisier period spectrum. We tentatively identified 
4 $\ell=2$ multiplets and 1 $\ell=1$ multiplet (see Table \ref{t2}), 
consistent with a rotation period of 18 hours. The rotational splitting 
analysis leaves 18 possible m=0 modes.

\section{Asteroseismological analysis}

The periods observed in white dwarfs are g-modes, where the restoring force is 
buoyancy. In a completely homogeneous star, g-modes would be evenly spaced in 
period. In reality, white dwarfs are differentiated and the chemical transition
zones induce a departure of the modes from their even spacings. The amount by 
which the modes deviate from their even spacing provides clues to the interior 
chemical structure of white dwarfs. In the presence of rotation, the modes
get evenly split in frequency. The magnitude of the frequency split depends
both on the rotational period and on the $\ell$ identification of the mode 
\citep{Unno89}.

A rotational frequency splitting of 7~\muhz ~ corresponds to a $\Delta m=1$ 
period split of 1 second for a 400~s mode and 8 seconds for a 1100~s mode. For 
$\ell$=2 modes, the corresponding frequency split is near 11~\muhz, 
corresponding to 2 and 13 seconds split in periods for the 400~s and 
1100~s modes respectively. These values are to compare with an average
period spacing of $\sim$42~s for $\ell$=1 modes and $\sim$27~s for 
$\ell$=2 modes. For asteroseismological fits, this means that 
lower radial overtone modes (lower k) are less sensitive to the exact m 
identification than higher k modes. Fitting lower k modes first therefore
minimizes modeling uncertainties due to our lack of knowledge of what the 
m identification of the modes is. On the other hand, lower k modes are
also more sensitive to core structure in the model and magnify our 
a priori ignorance of the core chemical profiles. As a first pass, we 
assumed that the observed modes were m=0 modes and looked for an 
asymptotic period spacing in the observed period spectra of G38-29 and R808.

For our asteroseismological fits, we used the White Dwarf Evolution code
(WDEC) to generate white dwarf models. The WDEC is described in detail 
in \citet{Lamb75} and \citet{Wood90}, and more recent modifications in 
\citet{Kim08a}. We used an iterative grid search method varying 6 parameters, 
listed in Table \ref{t1}, starting with a low resolution grid covering 
a wide region of parameter space in a range of effective temperature and 
mass suggested by the spectroscopic values. The grid is presented in 
Table \ref{t1}. Based on the fits found, we successively refined the 
grid, focusing on the regions of parameter space where the best fit 
models resided at each iteration. The final grids have a resolution of 50~K 
in effective temperature, 0.002~\msunm ~ in mass for G38-29, and 
0.005~\msunm ~ for R808, 0.002 (in the log) for the helium and hydrogen 
layer masses, 0.02 in central oxygen abundance and 0.02~$M_*$ for 
$X_{fm}$, the point where the oxygen abundance first starts to drop down.

\begin{table}[!ht]
\caption{
Parameters for the starting grid
\label{t1}
}
%\begin{center}
\begin{tabular}{llll}
\hline
Description & Parameter  & Range & Step size \\
\hline
Effective temperature  & \teffm & 10,600~-11,800~K     &  200~K        \\
Stellar mass           & \mstarm & 0.500~-~0.700~\msunm & 0.010~\msunm \\
Helium layer mass & \Mhem & $10^{-2.00}$~-~$10^{-2.40}$ & 0.20 in the log\\
Hydrogen layer mass    & \Mhm  & down to $10^{-7.00}$  & 0.20 in the log\\
Central oxygen abundance & \Xom & 0.00~-~1.00         & 0.10           \\
Point where oxygen abundance &&& \\
first drops to zero    & \qfmm  & 0.10~-0.80          & 0.10           \\
\hline
\end{tabular}
%\end{center}
\end{table}

\section{Results}
\label{results}

The results of the fits for G38-29 and R808 are presented in Table 
\ref{t2}. For G38-29, the best fit model has a temperature of 11,550~K, a
mass of 0.642~\msunm, a helium layer mass of $10^{-2.12}$, a hydrogen 
layer mass of $10^{-4.16}$. The central oxygen abundance is 0.80 and it 
starts dropping at mass point 0.68~$M_*$. For R808, the corresponding
parameters are 11,250~K, 0.675~\msunm, $10^{-2.58}$, $10^{-4.62}$, 
\Xom = 0.10, and $X_{fm}=0.84$.

\begin{table}[!ht]
\caption{
Mode identification and best fit model periods for G38-29 and R808
\label{t2}
}
%\begin{center}
\begin{tabular}{lcccrlcccr}
\hline
\multicolumn{5}{c}{G38-29} & \multicolumn{5}{c}{R808} \\
\hline
Observed & Model  & $\ell$ & k & m & Observed & Model  & $\ell$ & k & m \\
Period   & Period &        &   &   & Period   & Period &        &   &   \\
\hline
%\rule[-0.0cm]{0mm}{0.6cm}
432  & 428  & 1 &  7 &  0  & 405  & 406  & 1 & 6  &  0  \\
545  &      & 1 & 10 & +1  & 745  & 741  & 1 & 14 &  0  \\
547  & 545  & 1 & 10 &  0  & 875  & 870  & 1 & 17 &  0  \\
549  &      & 1 & 10 & -1  & 912  & 912  & 1 & 18 &  0  \\
706  & 705  & 1 & 14 &  0  & 915  &      & 1 & 18 &  -1 \\
709  &      & 1 & 14 & -1  &      &      &   &    &     \\
962  & 957  & 1 & 20 &  0  & 952  & 954  & 1 & 19 &  0  \\
1002 & 1002 & 1 & 21 &  0  & 1040 & 1042 & 1 & 21 &  0  \\
1082 &      & 1 & 23 &  0  &      &      &   &    &     \\
1089 & 1086 & 1 & 23 & +1  &      &      &   &    &     \\
\hline
413  & 409  & 2 & 14 &  0  &  511  & 514  & 2 & 17 &  0 \\
     &      &   &    &     &  629  &      & 2 & 22 & +1 \\
     &      &   &    &     &  632  & 637  & 2 & 22 &  0 \\
     &      &   &    &     &  796  & 788  & 2 & 28 &  0 \\
     &      &   &    &     &  843  & 837  & 2 & 30 &  0 \\
     &      &   &    &     &  860  & 862  & 2 & 31 &  0 \\
840  & 844  & 2 & 32 &  0  &  878  &      & 2 & 32 & +1 \\
     &      &   &    &     &  899  &      & 2 & 32 & -2 \\
     &      &   &    &     &  908  &      & 2 & 33 & +1 \\
     &      &   &    &     &  916  & 914  & 2 & 33 &  0 \\
     &      &   &    &     &  923  &      & 2 & 33 & -1 \\
900  & 900  & 2 & 34 &  0  &  952  &      & 2 & 35 & +1 \\
923  & 927  & 2 & 35 &  0  &  961  & 965  & 2 & 35 &  0 \\
945  & 940  & 2 & 36 &  0  &       &      &   &    &    \\ 
964  & 963  & 2 & 37 &  0  &  1011 & 1016 & 2 & 37 &  0 \\
980  &      & 2 & 38 & +1  &       &      &   &    &    \\
990  & 992  & 2 & 38 &  0  &  1042 & 1042 & 2 & 38 &  0 \\
1016 & 1022 & 2 & 39 &  0  &  1067 & 1067 & 2 & 39 &  0 \\
     &      &   &    &     &  1091 & 1092 & 2 & 40 &  0 \\
1087 & 1087 & 2 & 42 &  0  &  1144 & 1144 & 2 & 42 &  0 \\
\hline
\end{tabular}
%\end{center}
\end{table}

\begin{figure}[!h]
  \includegraphics[height=.46\textheight]{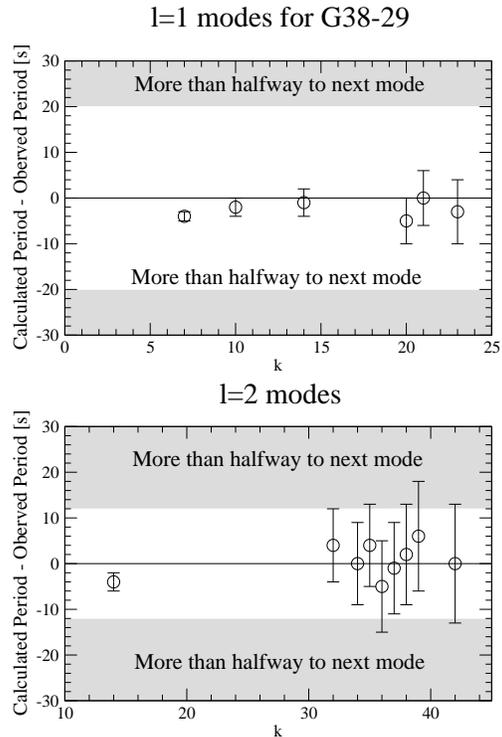}
  \caption{
  \label{f1}
  Quality of fit for the G38-29. The ``error bars" are not error bars, but 
  indicate the extent of the $\Delta m=\pm 1$ rotational period splitting 
  mentioned in the text. Note that this splitting is significantly
  less than the period spacing for $\ell=1$ modes, though they can lead to 
  significant uncertainties in the model periods and parameters if 
  misidentified in m value. For the $\ell=2$ modes, the frequency splitting
  translates to a period splitting that is of the same order as the average
  period spacing. The situation is aggravated by the fact that $m=\pm 2$ modes
  are allowed as well.}
\end{figure}

The quality of each period fit for G38-29 is illustrated in Figure \ref{f1}. 
For $\ell=1$ modes, the uncertain m identifications can lead to uncertainties 
in model parameters, but not in $\ell$ and k identification. For $\ell=2$ 
modes, the situation becomes more ambiguous for higher k modes, as the 
frequency splitting translates to a period splitting that is of the same order 
as the average period spacing. Consecutive multiplets start to overlap.

\section{Discussion}

In grid searches, the question often arises ``how fine is fine enough?''. How
finely do we need to sample parameter space in order to find any minimum 
present? Sampling tests are needed to fully answer that question, though we can
check that our solution approaches the minimum monotonically. The step sizes for
the final, high resolution grid quoted in the previous section appear 
sufficiently small. This also means that if one wanted to create a grid that 
covers the parameter ranges listed in Table \ref{t1}, one would have to 
calculate $\sim 10^{10}$ models. This is not manageable and the recourse is 
either to use ``smarter'' search algorithms such as genetic algorithms 
\citep{Metcalfe03b} or successively zooming in on the best fit. The two methods
are complementary to each other. Genetic algorithms, given enough modes, are 
effective at finding a global minimum while grid searches help refining the
minimum and provide more flexibility in trying different mode identifications.

The results presented here are preliminary, as we have made the assumption that
all observed modes that were not members of multiplets were m=0 modes. As 
noted earlier, this assumption does not significantly influence the fits for 
k$\leq$10 modes, but becomes important for higher k modes. Observationally, m=0
modes are not necessarily the higher amplitude modes in multiplets suggesting 
that single modes may not be m=0 modes. Lightcurve fitting techniques provide 
an independent way of mode identification \citep{Montgomery08}. It would be 
worth trying more general fits and using light curve fitting techniques for 
G38-29, where the observed frequencies are better determined.

\begin{theacknowledgments}
This research was supported by a Faculty Research grant from Georgia College \&
State University.
\end{theacknowledgments}

\bibliographystyle{aipproc}   % if natbib is available

\bibliography{agnes_kim}

\end{document}